\documentclass[prd,twocolumn,superscriptaddress,showpacs]{revtex4}
\usepackage[dvips]{graphicx}
\usepackage{color}
\usepackage{amsmath}
\usepackage{time}

\def\slash#1{#1\!\!\!\!\,/}

\begin{document}
%
%
%
\preprint{LAUR 05-8819}
\title[BCS]
   {Phases of a fermionic model with chiral condensates and Cooper pairs \\
   in 1+1 dimensions}
\author{Bogdan Mihaila}
\email{bmihaila@lanl.gov}
\affiliation{Theoretical Division,
   Los Alamos National Laboratory,
   Los Alamos, NM 87545}

\author{Krastan B. Blagoev}
\email{krastan@lanl.gov}
\affiliation{Theoretical Division,
   Los Alamos National Laboratory,
   Los Alamos, NM 87545}
\affiliation{The MIND Institute, Albuquerque, NM 87131}

\author{Fred Cooper}
\email{cooper@santafe.edu}
\affiliation{Santa Fe Institute,
   Santa Fe, NM 87501}
\affiliation{Theoretical Division,
   Los Alamos National Laboratory,
   Los Alamos, NM 87545}

\begin{abstract}
   We study the  phase structure of a 4-fermi model with three bare coupling
   constants, which potentially has three types of bound states. This
   model is a generalization of the model discussed previously by
   A.~Chodos~\emph{et~al.} [Phys. Rev. D~\textbf{61}, 045011
   (2000)], which contained both chiral condensates and Cooper
   pairs. For this generalization we find that there are two
   independent renormalized coupling constants which determine the
   phase structure at finite density and temperature.
   We find that the vacuum can be in one of three distinct phases depending
   on the value of these two renormalized coupling constants.
\end{abstract}



\maketitle


\section{Introduction}

Spontaneously broken symmetry phases occur in high-energy and
condensed matter physics by varying the temperature, density, or a
parameter controlling the interaction. These include chiral and
color broken symmetries in high-energy physics~(see e.g.
Refs.~\cite{raj,rapp,early}), as well as superconducting, and spin
and charge density waves ordered phases in condensed matter
systems~\cite{kbb1}. The phase diagram of these different physical
systems depends in general on the compatibility of the different
broken symmetries and usually is studied in the framework of the
Ginzburg-Landau-Wilson energy functional~\cite{kbb2}.

Recently we have studied the phase diagram of a relativistic model
with both chiral condensates and Cooper pairs~\cite{symmetric}, as
well as a non-relativistic model with ferromagnetic
superconductivity~\cite{kbb3}. For ferromagnetic superconductors it
is known~\cite{loff} that for spatially inhomogeneous order
parameters the ferromagnetic and superconducting phases coexist in a
small region of the phase diagram. In the context of relativistic
field theory this has been recently studied by Rajagopal's
group~\cite{raj2}.

In this paper we study a simple relativistic model in (1+1)
dimensions which combines the Gross-Neveu model~\cite{ref:GN} with a
model for Cooper pairs~\cite{ref:paper1}, plus an interaction which
splits the masses of the fermions. The symmetric version of this
model has been discussed previously~\cite{symmetric}, and exhibits
(at the mean-field level) a phase diagram which mimics some of the
features expected for QCD with two light flavors of quarks. The
model has a well defined $1/N$ expansion and is asymptotically free
so that it does not suffer from the cutoff dependencies of $3+1$
dimensional effective field theories considered by
others~\cite{rapp,ref:super1,ref:super2}. The question we are
addressing here is whether the splitting of the fermion masses can
lead to phase coexistence in 1+1 dimensions. What we find is that no
such phase coexistence occurs for static mean fields.

The presence of finite temperature condensates in one spatial
dimension violates the Mermin-Wagner theorem~\cite{ref:mermin}.
Nevertheless, Witten has argued~\cite{ref:Witten} that it is still
meaningful to study the formation of such condensates to leading
order in $1/N$, as the large-N expansion can give qualitatively good
understanding of the correlation functions, even when it gives the
wrong phase transition behavior. The quality of the approximation
should improve as we increase the number of spatial dimensions and
the mean-field critical behavior becomes exact above 3+1 dimensions.
Therefore, the 2-dimensional realization of the model presented in
this paper is a ``toy'' model exhibiting some of the features
expected to be true in 3+1 dimensions such as the restoration of the
symmetry at high temperatures.

The paper is organized as follows: In Sec.~\ref{model} we discuss
the model and derive the (1+1) dimensional effective potential in
the leading-order large $N$ approximation. The renormalization of
the effective potential is performed in Sec.~\ref{renorm}. We
discuss the possible phases of the vacuum state in
Sec.~\ref{vacuum}, and study the phase structure of our
2-dimensional model in Sec.~\ref{phase}. Our conclusions are
presented in Sec.~\ref{concl}.


\section{Model}
\label{model}

We would like to make the simplest generalization of our previous
model~\cite{symmetric} which has $N$ families of a two-flavor model,
but also contains an interaction that treats the two flavors
differently. The flavors can be thought of as up and down quarks, or
proton and neutron. Thus we are lead to consider a model described
by the 2N -component Lagrangian
\begin{align}
   \mathcal{L}  \ = & \
   \bar \psi^{(i,\, a)}
   \bigl ( \mathrm{i} \slash{\partial} - \mu \gamma^0 \bigr ) \psi^{(i,\, a)}
   \\ \notag &
   +
   \frac{g_1^2}{2}
   [\bar \psi^{(i)} \psi^{(i,\, a)}]
   [\bar \psi^{(j)} \psi^{(j,\, a)}]
   +
   \frac{g_2^2}{2}
   [\bar \psi^{(i,\, a)} \tau_{3, ij} \psi^{(j,\, a)}]^2
   \\ \notag &
   -
   G^2 \,
   [\epsilon_{\alpha \beta} \psi^{(i,\, a)\dag}_\alpha \psi^{(i,\, a)\dag}_\beta]
   [\epsilon_{\gamma \delta} \psi^{(j,\, a)}_\gamma \psi^{(j,\, a)}_\delta]
   \>.
\end{align}
The superscripts indicate the flavor indices, $i=1,2$ and the family
index $a=1,2,\ldots,n$ . We will treat this model in the leading
order in large-N approximation which is a mean-field approximation.
The coupling constants $g_i^2$ and $G^2$ must generically scale as
$\lambda/N$, with $\lambda$ fixed, in order to obtain the large $N$
limit. For simplicity, in the following we drop the index $a$.

To perform the Hubbard-Stratonovich transformation~\cite{ref:hub},
we first introduce auxiliary fields by adding to the Lagrangian the
terms
\begin{align}
   \Delta \mathcal{L}
   \ = \ &
   - \frac{1}{2 g_1^2}
   \Bigl \{
   m_1 + g_1^2
   [\bar \psi^{(i)} \psi^{(i)}]
   \Bigr \}^2
   \\ \notag &
   - \frac{1}{2 g_2^2}
   \Bigl \{
   m_2 + g_2^2
   [\bar \psi^{(i)} \tau_{3, ij} \psi^{(j)}]
   \Bigr \}^2
   \\ \notag &
   -
   \frac{1}{G^2} \,
   [B^\dag - G^2 \epsilon_{\alpha \beta} \bar \psi^{(i)}_\alpha \psi^{(i)}_\beta]
   [B + G^2 \epsilon_{\gamma \delta} \bar \psi^{(j)}_\gamma \psi^{(j)}_\delta]
   \>.
\end{align}
In $\mathcal{L}' = \mathcal{L} + \Delta \mathcal{L}$, the terms
quartic in the fermion fields cancel, and we can formally write
\begin{align}
   \mathcal{L}'  \ = \ &
   - \frac{m_1^2}{2 g_1^2}
   - \frac{m_2^2}{2 g_2^2}
   - \frac{B^\dag B}{G^2}
   \\ \notag &
   - \frac{1}{2}
   \bigl [
   \psi^\dag \psi \bigr ]
\left [
\begin{array}{cc}
   - h & 2 \, (\epsilon B^\dag) \\
   - 2 \, (\epsilon B) & h^T
\end{array} \right ]
\left [
\begin{array}{c}
   \psi \\
   \psi^\dag
\end{array} \right ]
   \>,
\end{align}
where $\epsilon = -\mathrm{i}\, \sigma_2$, and
$
   h =
   \gamma^0 \bigl ( \mathrm{i} \slash{\partial} - \mu \gamma^0 - m_1 \bigr )
   - m_2 \gamma^0 \tau_3
$. We integrate out $\psi$ and $\psi^\dag$ and realize that the
resulting action is proportional to $N$. Performing  the integration
over the auxiliary fields by steepest descent and Legendre
transforming the generating functional we obtain, in the standard
manner,  the leading order in large-N effective action
\begin{align}
   & \Gamma_{\mathrm{eff}} (m_1, m_2, M)
   \\ \notag &
   =
   \int \mathrm{d}^d x \,
   \Bigl (
   - \frac{m_1^2}{2 g_1^2}
   - \frac{m_2^2}{2 g_2^2}
   - \frac{M^2}{4G^2} \Bigr )
   - V^{(1)}(m_1,m_2,M)
   \>,
\end{align}
with $M^2=4 \, B^\dag B$, $\tilde h = \sigma_2 h \sigma_2$,
and~\cite{lemma}
\begin{equation}
   V^{(1)}=\frac{\mathrm{i}}{2} \,
   \mathrm{Tr} \ln ( h^T h )
   + \frac{\mathrm{i}}{2} \,
   \mathrm{Tr} \ln \bigl [ 1 + M^2/(\tilde h h^T) \bigr ]
   \>.
\end{equation}
The gap equations are obtained by setting to zero the derivatives
of the effective action with respect to~$m_i^2$ and~$M^2$,
i.e.~\cite{1std}
\begin{align}
   \delta_{m_i^2} \Gamma
   = 0 = & - \frac{1}{2 g_i^2}
   - \frac{\mathrm{i}}{2}
   \int [ \mathrm{d}^d k ] \
   \Bigl \{
      \mathrm{Tr} \, \Bigl [ \frac{1}{h^T h} \frac{\delta(h^T h)}{\delta m_i^2}
                     \Bigr ]
   \\ \notag & \qquad
      -
      M^2 \mathrm{Tr} \,
      \Bigl [ \frac{1}{M^2 + \tilde h h^T}
              \frac{1}{\tilde h h^T} \frac{\delta(\tilde h h^T)}{\delta m_i^2}
      \Bigr ]
   \Bigr \}
   \>,
\\
   \delta_{M^2} \Gamma
   = 0 = & - \frac{1}{4G^2}
   - \frac{\mathrm{i}}{2}
   \int [ \mathrm{d}^d k ] \
      \mathrm{Tr} \,
      \frac{1}{M^2 + \tilde h h^T}
   \>,
\end{align}
with $[\mathrm{d}^d k] = \mathrm{d}^d k /(2\pi)^d$. The effective
potential is obtained as
\begin{align}
   V_{\mathrm{eff}}(m_i,M)
   \!= &
   \frac{m_1^2}{2 g_1^2}
   + \! \frac{m_2^2}{2 g_2^2}
   + \! \frac{M^2}{4G^2}
   + \!V^{(1)}(m_1,m_2,M)
   \>.
\end{align}



In (1+1) dimensions, a convenient representation for $\gamma^\mu$ is
given by the Pauli matrices, as $\gamma^0 = \sigma_1$ and $\gamma^1
= - i \sigma_2$. This gives
\begin{align}
   h \ = \ &
   (k_0 - \mu) - m_1 \sigma_1 - k_1 \sigma_3
   - m_2 \sigma_1 \tau_3
   \>,
   \\
   h^T \ = \ &
   - (k_0 + \mu) - m_1 \sigma_1 + k_1 \sigma_3
   - m_2 \sigma_1 \tau_3
   \>.
\end{align}
The pure Cooper pair version of this model, $m_1=m_2=0$, has been
discussed in Ref.~\cite{ref:paper1}, while the symmetric version
($m_2=0$) was studied in Ref.~\cite{symmetric}. Here, we follow
closely the approach outlined in these references.


\subsection{symmetric case: $m_2=0$}

In the symmetric case~\cite{symmetric}, the effective potential is
isospin independent, and the isospin \emph{trace} results in a
multiplicative  factor of~2.

Considering first the zero temperature case we introduce the
notations: $h^T h = A + \vec B \cdot \vec \sigma$ and $\tilde h h^T
= A' + \vec B' \cdot \vec \sigma$. Here, we have~\cite{1+1s}: $A^2 -
\vec B \cdot \vec B = A'^2 - \vec B' \cdot \vec B'$. Hence, we
obtain the derivatives
\begin{align}
   \delta_{m_1^2} V^{(1)}
   = & \frac{1}{m_1}
   \frac{(A' + M^2) \delta_{m_1^2} A' - \vec B' \cdot \delta_{m_1^2} \vec B'}
        {(k_0^2 - k_+^2)(k_0^2 - k_-^2)}
   \>,
   \\
   \delta_{M^2} V^{(1)}
   = &
   \frac{2(A' + M^2)}
        {(k_0^2 - k_+^2)(k_0^2 - k_-^2)}
   \>,
\end{align}
and the gap equations are
\begin{align}
   \frac{1}{2g_1^2}
   & = \mathrm{i} \int [\mathrm d^2 k] \
   \frac{f_{m_1^2}(k_0,k_1;m_1)}{(k_0^2 - k_+^2)(k_0^2 - k_-^2)}
   \>,
   \\
   \frac{1}{4G^2}
   & = \mathrm{i} \int [\mathrm d^2 k] \
   \frac{f_{M^2}(k_0,k_1;m_1)}{(k_0^2 - k_+^2)(k_0^2 - k_-^2)}
   \>,
\end{align}
where $[\mathrm{d}^d k] = \mathrm{d}^d k / (2\pi)^d$, and
$k_\pm^2(m_1) = a(m_1) \pm 2 b(m_1)$, with
\begin{align*}
   a(m) & = k_1^2 + \mu^2 + M^2 + m^2
   \>,
   \\
   b(m) & = \bigl [ \mu^2 k_1^2 + (\mu^2 + M^2) m^2 \bigr ]^{1/2}
   \>.
\end{align*}
We also have
\begin{align*}
   f_{m_1^2}(k_0,k_1;m) & =
   k_0^2 - k_1^2 + (\mu^2 + M^2 - m^2)
   \>,
   \\
   f_{M^2}(k_0,k_1;m) & =
   k_0^2 - k_1^2 - (\mu^2 + M^2 - m^2)
   \>.
\end{align*}
Next, we perform the $k_0$ integral. Generically, we have
\begin{align}
   &
   \int_{-\Lambda}^{\Lambda} [\mathrm d k_1]
   \int_{-\infty}^{\infty} [\mathrm d k_0] \
   \frac{f(k_0,k_1;m)}{(k_0^2 - k_+^2)(k_0^2 - k_-^2)}
   \\ \notag &
   =
   - \mathrm{i}
   \int_{-\Lambda}^{\Lambda} [\mathrm d k_1] \
   \frac{1}{4 b}
   \Bigl [
   \frac{f(k_+,k_1;m)}{k_+} - \frac{f(k_-,k_1;m)}{k_-} \Bigr ]
\end{align}
Hence, the gap equations become
\begin{align*}
   \frac{1}{2g_1^2}
   & =
   \frac{1}{4}
   \int_{-\Lambda}^{\Lambda} [\mathrm d k_1]
   \Bigl [
   \frac{1}{k_+} + \frac{1}{k_-}
   + \frac{\mu^2 + M^2}{b}
   \Bigl ( \frac{1}{k_+} - \frac{1}{k_-} \Bigr )
   \Bigr ]
   \>,
   \\
   \frac{1}{4G^2}
   & =
   \frac{1}{4}
   \int_{-\Lambda}^{\Lambda} [\mathrm d k_1]
   \Bigl [
   \frac{1}{k_+} + \frac{1}{k_-}
   + \frac{m_1^2}{b}
   \Bigl ( \frac{1}{k_+} - \frac{1}{k_-} \Bigr )
   \Bigr ]
   \>.
\end{align*}
The $k_1$ integrals are logarithmically divergent and need to be
regularized by imposing a cutoff $\Lambda$. The gap equations can be
obtained by direct differentiation of the effective potential,
$V_{\mathrm{eff}}(m_1,0,M)$, corresponding to
\begin{align}
   V^{(1)}(m_1,0,M) =
   - \!\!
   \int_{0}^{\Lambda} [\mathrm d k_1] \,
   \Bigl [ k_+(m_1) + k_-(m_1) \Bigr ]
   \>.
\end{align}

The finite temperature gap equations can be obtained formally from
the zero temperature ones, with the replacements
\begin{align}
   \frac{1}{k_+} & \rightarrow
   \frac{1}{k_+} [ 1 - 2 n_F(k_+) ]
   \>,
   \notag \\
   \frac{1}{k_-} & \rightarrow
   \frac{1}{k_-} [ 1 - 2 n_F(k_-) ]
   \>,
\end{align}
where $n_F(k)=[e^{\beta k}+1]^{-1}$ is the Fermi-Dirac distribution
function, with $\beta=1/T$. It follows that the finite temperature
effective potential corresponds to
\begin{align}
   V^{(1)}&(m_1,0,M) =
   - \!\!
   \int_{0}^{\Lambda} [\mathrm d k_1] \,
   \Bigl \{
   k_+(m_1) + k_-(m_1)
   \\ \notag &
   + \frac{2}{\beta} \ln\bigl [ 1+e^{-\beta k_+}(m_1) \bigr ]
   + \frac{2}{\beta} \ln\bigl [ 1+e^{-\beta k_-}(m_1) \bigr ]
   \Bigr \}
   \>.
\end{align}


\subsection{asymmetric case: $m_2\neq0$}

We have~\cite{1+1a}: $h^T h = A + \vec B \cdot \vec \sigma + ( C +
\vec D\cdot \vec \sigma) \tau_3$ and $\tilde h h^T = A' + \vec B'
\cdot \vec \sigma + ( C' + \vec D' \cdot \vec \sigma) \tau_3$. It is
convenient to use an explicit isospin matrix representation, i.e.
\begin{align*}
   h^T h & =
   \Bigl [
   \begin{array}{cc}
   A + C + (\vec B + \vec D) \cdot \sigma & 0 \\
   0 & A - C + (\vec B - \vec D) \cdot \sigma
   \end{array}
   \Bigr ]
   \>,
   \\
   \tilde h h^T & =
   \Bigl [
   \begin{array}{cc}
   A' \!+ C' \!\!+ (\vec B' \!+ \vec D') \cdot \sigma & 0 \\
   0 & A' \!- C' \!\!+ (\vec B' \!- \vec D') \cdot \sigma
   \end{array}
   \Bigr ]
   \>.
\end{align*}
By inspection, in the asymmetric case we find that the isospin trace
gives two contributions similar to the symmetric case, corresponding
to masses $m_{\pm}=m_1\pm m_2$.
We have
\begin{widetext}
\begin{align}
   V^{(1)} & 
   \equiv
   -
   \frac{1}{2}
   \int_{0}^{\Lambda} [\mathrm d k_1]
   \Bigl \{
   k_+(m_+)
   + k_-(m_+)
   + k_+(m_-)
   + k_-(m_-)
   \\ \notag &
   + \frac{2}{\beta} \ln\bigl [ 1\!+ \!e^{-\beta k_+(m_+)}\bigr ]
   \!+ \! \frac{2}{\beta} \ln\bigl [ 1\!+ \!e^{-\beta k_-(m_+)}\bigr ]
   \!+ \!\frac{2}{\beta} \ln\bigl [ 1\!+ \!e^{-\beta k_+(m_-)}\bigr ]
   \!+ \! \frac{2}{\beta} \ln\bigl [ 1\!+ \!e^{-\beta k_-(m_-)}\bigr ]
   \Bigr \}
   \>.
\end{align}


\section{Renormalized effective potential}
\label{renorm}

It is sufficient to perform renormalization at zero temperature and
chemical potential ($T=\mu=0$). In this context, we can define the
renormalized coupling constant in terms of the physical scattering
amplitude of fermions at a particular momentum scale. The addition
of $\mu$ and $T$ will only result in finite corrections to the gap
equations, and therefore to the vacuum values of $m_i$ and~$M$.


\subsection{$T=\mu=0$}

In order to expose the divergences in the $k_1$ integral, we
consider now the integral
\begin{align*}
   \int \! \mathrm{d}k \sqrt{k^2 + m^2} \!=\!
   \frac{k}{2} \sqrt{k^2 + m^2}
   + \frac{m^2}{2} \ln \bigl ( k + \!\sqrt{k^2 + m^2} \bigr )
   \>,
\end{align*}
which indicates that we have both logarithmic and quadratic
ultraviolet divergences. First, we eliminate the quadratic
divergence by adding a constant term at a fixed momentum
cutoff,~$\Lambda$, i.e.
\begin{align}
   V^{(1)} (m_1, m_2, & M) \!\rightarrow
   -
   \frac{1}{2} \!
   \int_{0}^{\Lambda} \![\mathrm d k_1]
   \Bigl [
   k_+(m_-)
   + k_-(m_-)
   + k_+(m_+)
   + k_-(m_+)
   - 4 k_1
   \Bigr ]
   \>.
\end{align}
One more integral~\cite{int1} yields the unrenormalized effective
potential at $T=\mu=0$, as
\begin{align}
   & V_{\mathrm{eff}}(m_i,M) = \
   M^2 \,
   \Bigl ( \frac{1}{4G^2} - \frac{1}{4\pi} \Bigr )
   +
   m_1^2 \,
   \Bigl ( \frac{1}{2g_1^2} - \frac{1}{4\pi} \Bigr )
   +
   m_2^2 \,
   \Bigl ( \frac{1}{2g_2^2} - \frac{1}{4\pi} \Bigr )
\label{veff_a2}
   \\ \notag &
   - \frac{1}{8\pi}
   \biggl [
   (M+m_+)^2 \ln \frac{2\Lambda}{|M+m_+|}
   \!+\!
   (M-m_+)^2 \ln \frac{2\Lambda}{|M-m_+|}
   \!+\!
   (M+m_-)^2 \ln \frac{2\Lambda}{|M+m_-|}
   \!+\!
   (M-m_-)^2 \ln \frac{2\Lambda}{|M-m_-|}
   \biggr ]
   \>.
\end{align}
Noting that $2(M^2 + m_1^2 + m_2^2) = (M^2 + m_+^2) + (M^2 +
m_-^2)$, we renormalize by requiring that the renormalized coupling
constants, $g_i^2$ and $G^2$, satisfy~\cite{2ndd}
\begin{align}
   \delta^2_{m_i^2} V_{\mathrm{eff}} \!\bigr |_{m_{i0},M_0}
   \!= \frac{1}{g_{i,\mathrm{R}}^2}
   \>, \quad
   \delta_{B^\dag} \delta_{B}  V_{\mathrm{eff}} \!\bigr |_{m_{i0},M_0}
   \!= \frac{1}{G_{\mathrm{R}}^2}
   \>.
\label{eq:1+1_ren}
\end{align}
Here, the masses $\{m_{i0},M_0\}$ have arbitrary renormalization
values on which the coupling constants will depend.
Eqs.~\eqref{eq:1+1_ren} are solved for $g_i^2$ and $G^2$ as a
function of $g_{i,\mathrm{R}}^2$ and $G_{\mathrm{R}}^2$. We obtain
\begin{align}
   \frac{1}{2g_i^2} = \ &
   \frac{1}{2g_{i\mathrm{R}}^2} - \frac{1}{2\pi}
   - \frac{1}{4\pi} \ln \frac{\gamma}{(2\Lambda)^2}
   \>,
\label{eq:giR}
\end{align}
and
\begin{align}
   &
   \frac{1}{4G^2} =
   \frac{1}{4G_\mathrm{R}^2} - \!\frac{1}{4\pi}
   - \!\frac{1}{4\pi} \ln \frac{\gamma}{(2\Lambda)^2}
\label{eq:GR}
             + \frac{1}{16\pi}\frac{m_{+0}}{M_0} \,
             \ln \Bigl | \frac{M_0 - m_{+0}}{M_0 + m_{+0}} \Bigr |
             + \frac{1}{16\pi}\frac{m_{-0}}{M_0} \,
             \ln \Bigl | \frac{M_0 - m_{-0}}{M_0 + m_{-0}} \Bigr |
   \>,
\end{align}
where $m_{\pm 0} = m_{10} \pm m_{20}$, and we have introduced the
renormalization scale
\begin{align}
   \gamma^2 = & (M_0^2-m_{10}^2-m_{20}^2)^2 - 4 m_{10}^2 m_{20}^2
   =
   (M_0^2-m_{+0}^2)(M_0^2-m_{-0}^2)
\label{gamma0}
   \>.
\end{align}
Hence, the renormalized effective potential at $T=\mu=0$ is obtained
by substituting in Eq.~\eqref{veff_a2}, the \emph{bare} coupling
terms as given by Eqs.~\eqref{eq:giR} and~\eqref{eq:GR}. We obtain
\begin{align}
   V_{\mathrm{eff}}(m_i, M) = \ &
   M^2 \beta + m_1^2 \alpha_1 + m_2^2 \alpha_2
   \!+\! \frac{M^2+m_+^2}{8\pi}
                   \ln \frac{\bigl | M^2 \!- m_+^2 \bigr |}
                       {\gamma}
   \!+\! \frac{M^2+m_-^2}{8\pi}
                   \ln \frac{\bigl | M^2 \!- m_-^2 \bigr |}
                       {\gamma}
   \notag \\ &
      \!-\! \frac{M m_+}{4\pi} \ln \Bigl | \frac{M-m_+}
                                            {M+m_+} \Bigr |
      \!-\! \frac{M m_-}{4\pi} \ln \Bigl | \frac{M-m_-}
                                            {M+m_-} \Bigr |
   \>,
\end{align}
with
\begin{align}
   \alpha_i = \ & \frac{1}{2g_{i\mathrm{R}}^2} - \frac{3}{4\pi}
   \>,
   \\
   \beta = \ & \frac{1}{4G_\mathrm{R}^2} - \frac{1}{2\pi}
             + \frac{1}{16\pi}\frac{m_{+0}}{M_0}
             \ln \Bigl | \frac{M_0 - m_{+0}}{M_0 + m_{+0}} \Bigr |
             + \frac{1}{16\pi}\frac{m_{-0}}{M_0}
             \ln \Bigl | \frac{M_0 - m_{-0}}{M_0 + m_{-0}} \Bigr |
   \>.
\end{align}
Correspondingly, the gap equations are
\begin{align}
   & m_+ \!
   \Bigl ( \frac{\bar \alpha_1}{2}
      \!+\! \frac{1}{8\pi} \ln \! \frac{|M^2 - m_+^2|}{\gamma}
   \Bigr )
   \!+\! m_- \!
   \Bigl ( \frac{\bar \alpha_1}{2}
     \!+\! \frac{1}{8\pi} \ln \! \frac{|M^2 - m_-^2|}{\gamma}
   \Bigr )
   \!-\! \frac{M}{8\pi} \!
     \Bigl (
        \ln \! \Bigl | \! \frac{M - m_+}{M + m_+} \! \Bigr |
        \!+\!
        \ln \! \Bigl | \! \frac{M - m_-}{M + m_-} \! \Bigr |
     \Bigr )
   \! = \! 0
   \>,
   \\
   & m_+ \!
   \Bigl ( \frac{\bar \alpha_2}{2}
      \!+\! \frac{1}{8\pi} \ln \! \frac{|M^2 - m_+^2|}{\gamma}
   \Bigr )
   \!-\! m_- \!
   \Bigl ( \frac{\bar \alpha_2}{2}
     \!+\! \frac{1}{8\pi} \ln \! \frac{|M^2 - m_-^2|}{\gamma}
   \Bigr )
   \!-\! \frac{M}{8\pi} \!
     \Bigl (
        \ln \! \Bigl | \! \frac{M - m_+}{M + m_+} \! \Bigr |
        \!-\!
        \ln \! \Bigl | \! \frac{M - m_-}{M + m_-} \! \Bigr |
     \Bigr )
   \! = \! 0
   \>,
   \\
   & M \!
   \Bigl ( \bar \beta
      \!+\! \frac{1}{8\pi}
        \ln \frac{|M^2 - m_+^2|}{\gamma}
      \!+\! \frac{1}{8\pi}
        \ln \frac{|M^2 - m_-^2|}{\gamma}
     \Bigr )
   \!-\! \frac{m_+}{8\pi} \ln \Bigl | \frac{M-m_+}{M+m_+} \Bigr |
   \!-\! \frac{m_-}{8\pi} \ln \Bigl | \frac{M-m_-}{M+m_-} \Bigr |
   \! = \! 0
   \>,
\end{align}
with
\begin{equation}
   \bar \alpha_i = \alpha_i + \frac{1}{4\pi}
   \>, \qquad
   \bar \beta = \beta + \frac{1}{4\pi}
   \>.
\end{equation}
The solutions $m_i=m_i^*$ and $M=M^*$ give the local extrema of
$V_{\mathrm{eff}}$, and represent physical parameters that must be
independent of the renormalization scale $\gamma$. We note that if
we solve for the combinations $\delta_i = \beta - \alpha_i$, the
renormalization scale $\gamma$ drops out. Therefore, $\delta_i$ are
true physical parameters in the theory, and their values control
which of the condensates $m_i$ and $M$ can exist. The third physical
parameter in this model is $\Delta \delta = \delta_2 - \delta_1 =
\alpha_1 - \alpha_2$, which is also independent of~$\gamma$. Of
course, only two of the three physical parameters $\delta_1$,
$\delta_2$ and $\Delta \delta$, are independent of each other.


\subsection{Finite $T$ and $\mu$}
\label{Veff}

As advertised, the subtractions necessary to remove the ultraviolet
divergences at $T=\mu=0$, are sufficient to renormalize the
effective potential at finite temperature and chemical potential. In
order to see this, we note that we can write the bare coupling
constants~\eqref{eq:giR} and~\eqref{eq:GR} as
\begin{align}
   \frac{1}{2g_i^2} = \
   \bar \alpha_i + X
   \>, \quad
   \frac{1}{4G^2} = \
   \bar \beta + X
   \>,
\end{align}
where $X$ is the divergent integral
\begin{align}
   X = &
   \frac{1}{4} \int_{0}^{\Lambda} [ \mathrm{d}k_1 ]
   \biggl [ \frac{1}{\sqrt{k_1^2 + (M_0+m_{+0})^2}} \!+\! \frac{1}{\sqrt{k_1^2 + (M_0-m_{+0})^2}}
   \!+\! \frac{1}{\sqrt{k_1^2 + (M_0+m_{-0})^2}} \!+\! \frac{1}{\sqrt{k_1^2 + (M_0-m_{-0})^2}}
   \biggr ]
   \notag \\ = &
   \frac{1}{4\pi} \ln \frac{(2\Lambda)^2}{\gamma}
   + \mathrm{terms \ that \ vanish \ as \ \Lambda \rightarrow \infty}
   \>.
\end{align}
Then, the full renormalized effective potential at finite
temperature and chemical potential can be written as
\begin{align}
   & V_{\mathrm{eff}}(m_i,M) =
   M^2 \bar \beta
   \!+\! m_1^2 \bar \alpha_1 \!+\! m_2^2 \bar \alpha_2
   \!-\!
   \frac{1}{2} \!
   \int_{0}^{\infty} \!\! [\mathrm d k_1]
   \biggl \{
   k_+(m_+)
   \!+\! k_-(m_+)
   \!+\! k_+(m_-)
   \!+\! k_-(m_-)
   \!-\! 4 k_1
\label{finite}
   \\ & \notag
   +\! \frac{2}{\beta} \ln\bigl [ 1\!+ \!e^{-\beta k_+(m_+)}\bigr ]
   \!+\! \frac{2}{\beta} \ln\bigl [ 1\!+ \!e^{-\beta k_-(m_+)}\bigr ]
   \!+\! \frac{2}{\beta} \ln\bigl [ 1\!+ \!e^{-\beta k_+(m_-)}\bigr ]
   \!+\! \frac{2}{\beta} \ln\bigl [ 1\!+ \!e^{-\beta k_-(m_-)}\bigr ]
   \\ \notag &
   -\! \frac{M^2 \!+m_1^2 \!+m_2^2}{2} \!
   \biggl [ \frac{1}{\sqrt{k_1^2 + (M_0+m_{+0})^2}}
   \!+\! \frac{1}{\sqrt{k_1^2 + (M_0-m_{+0})^2}}
   \!+\! \frac{1}{\sqrt{k_1^2 + (M_0+m_{-0})^2}}
   \!+\! \frac{1}{\sqrt{k_1^2 + (M_0-m_{-0})^2}}
   \biggr ]
   \biggr \}
   \>.
\end{align}
\end{widetext}


\section{Vacuum}
\label{vacuum}

\begin{figure}[t]
   \includegraphics[width=0.8\columnwidth]{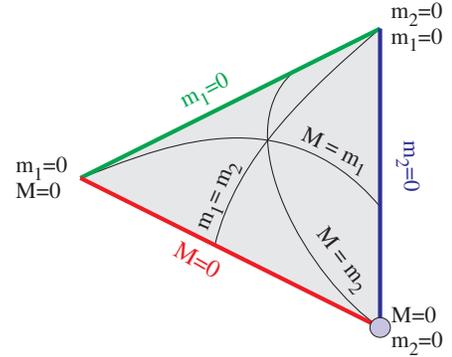}
   \caption{\label{triangle}(Color online)
   Parameter space for the effective potential at $T=\mu=0$.
   The corner with $M=m_2=0$ indicates the QCD-like sector.}
\end{figure}

At $T=\mu=0$, the minimum effective potential takes the form
\begin{align}
   V_{\mathrm{eff}}(m_i,M) = &
   - \frac{1}{4\pi}(M^2+m_1^2+m_2^2)
   \\ \notag
   = &
   - \frac{1}{8\pi}(M^2+m_+^2)
   - \frac{1}{8\pi}(M^2+m_-^2)
   \>,
\end{align}
where, for simplicity, we have dropped the $^*$ notation of the
parameter values at the minimum. We analyze now the solutions of the
gap equations and determine the true vacuum of the theory as the
\emph{global} minimum of the effective potential. The parameter
space corresponds to the triangle depicted in Fig.~\ref{triangle}.
Here, the corners of the triangle correspond to the situation when
two masses are zero:
\begin{itemize}
   \item $m_1=m_2=0$: $M^2=\gamma\, e^{-(1+4\pi\beta)}$,
   and
\begin{equation}
   V_{\mathrm{eff}}(0,0,M) =
   - \frac{\gamma}{4\pi} \ e^{-(1+4\pi\beta)}
   \>,
\end{equation}

   \item $M=m_2=0$: $m_1^2=\gamma\, e^{-(1+4\pi\alpha_1)}$,
   and
\begin{equation}
   V_{\mathrm{eff}}(m_1,0,0) =
   - \frac{\gamma}{4\pi} \ e^{-(1+4\pi\alpha_1)}
   \>,
\end{equation}

   \item $m_1=M=0$: $m_2^2=\gamma\, e^{-(1+4\pi\alpha_2)}$,
   and
\begin{equation}
   V_{\mathrm{eff}}(0,m_2,0) =
   - \frac{\gamma}{4\pi} \ e^{-(1+4\pi\alpha_2)}
   \>,
\end{equation}

\end{itemize}
while the sides of the triangle correspond to the case where one
mass is zero. We have

\begin{itemize}
   \item $M=0$. Here, we have $m_+ \neq m_-$ $(m_2
\neq 0)$, and the gap equations become
\begin{align}
   m_1
   \Bigl (
   \bar \alpha_1
   + \frac{1}{4\pi} \ln \frac{|m_2^2 - m_1^2|}{\gamma}
   \Bigr )
   - \frac{m_2}{4\pi} \ln \Bigl | \frac{m_1 - m_2}{m_1 + m_2} \Bigr |
   & = 0
   \>,
\label{eq:m1f_gap}
   \\
   m_2
   \Bigl (
   \bar \alpha_2
   + \frac{1}{4\pi} \ln \frac{|m_2^2 - m_1^2|}{\gamma}
   \Bigr )
   - \frac{m_1}{4\pi} \ln \Bigl | \frac{m_1 - m_2}{m_1 + m_2} \Bigr |
   & = 0
   \>.
\label{eq:m2f_gap}
\end{align}

   \item $m_1=0$. In this case, we have $m_+ = - m_- = m_2$.
The gap equations lead to
\begin{align}
   m_2
   \Bigl ( \bar \alpha_2
      + \frac{1}{4\pi} \ln \frac{|M^2 - m_2^2|}{\gamma}
   \Bigr )
   - \frac{M}{4\pi}
        \ln \Bigl | \frac{M - m_2}{M + m_2} \Bigr |
   & = 0
   \>,
   \\
   M
   \Bigl ( \bar \beta
      + \frac{1}{4\pi}
        \ln \frac{|M^2 - m_2^2|}{\gamma}
     \Bigr )
   - \frac{m_2}{8\pi} \ln \Bigl | \frac{M-m_2}{M+m_2} \Bigr |
   & = 0
   \>.
\end{align}

   \item $m_2=0$ and we recover the
symmetric-case results with $( m_1 \neq 0,\, M \neq 0 )$:
\begin{align}
   & m_1
   \Bigl ( \bar \alpha_1
      + \frac{1}{4\pi} \ln \frac{|M^2 - m_1^2|}{\gamma}
   \Bigr )
   \!-\! \frac{M}{4\pi}
                   \ln \Bigl | \frac{M - m_1}
                                    {M + m_1} \Bigr |
   \!= 0
   \>,
   \\
   & M
   \Bigl ( \bar \beta
      + \frac{1}{4\pi} \ln \frac{|M^2 - m_1^2|}{\gamma}
   \Bigr )
      - \frac{m_1}{4\pi}
                   \ln \Bigl | \frac{M - m_1}
                                    {M + m_1} \Bigr |
   \!= 0
   \>.
\end{align}
\end{itemize}

\begin{table}[t]
\caption{\label{corners}Position of the global minimum of the
effective potential at $T=\mu=0$.}
\begin{ruledtabular}
\begin{tabular}{lcr}
Parameters & 
             masses & $V_{\mathrm{eff}}(m_i,M)$
\\
\colrule $\delta_1<\delta_2<0$     $(\Delta \delta > 0)$ &
                          $M  >m_1>m_2$ & $V_{\mathrm{eff}}(0,0,M)$ \\
$\delta_2<\delta_1<0$     $(\Delta \delta < 0)$ &
                          $M  >m_2>m_1$ & $V_{\mathrm{eff}}(0,0,M)$ \\
$\delta_1<0<\delta_2$     $(\Delta \delta > 0)$ &
                          $m_2>M  >m_1$ & $V_{\mathrm{eff}}(0,m_2,0)$ \\
$0<\delta_1<\delta_2$     $(\Delta \delta > 0)$ &
                          $m_2>m_1>M  $ & $V_{\mathrm{eff}}(0,m_2,0)$ \\
$0<\delta_2<\delta_1$     $(\Delta \delta < 0)$ &
                          $m_1>m_2>M  $ & $V_{\mathrm{eff}}(m_1,0,0)$ \\
$\delta_2<0<\delta_1$     $(\Delta \delta < 0)$ &
                          $m_1>M  >m_2$ & $V_{\mathrm{eff}}(m_1,0,0)$
\end{tabular}
\end{ruledtabular}
\end{table}

When only one mass ($m_1$, $m_2$, or $M$) is zero, (see appendix of
Ref.~\cite{symmetric}) the ``global'' minimum, along the side of the
triangle, corresponds to one of its ends and which end exactly has
the lower potential depends on the value of the parameters
$\delta_1$, $\delta_2$, or $\Delta \delta$, respectively:
\begin{itemize}

   \item If $\delta_1<0$, then $V_{\mathrm{eff}}(0,0,M)$ is the ``global''
minimum along the \underline{$m_2=0$ line}, whereas if $\delta_1>0$
then the ``global'' minimum of the effective potential is
$V_{\mathrm{eff}}(m_1,0,0)$. The critical value $\delta_1=0$
corresponds to $M=m_1$.

    \item If $\delta_2<0$, then $V_{\mathrm{eff}}(0,0,M)$ is the ``global''
minimum along the \underline{$m_1=0$ line}, whereas if $\delta_2>0$
then the ``global'' minimum of the effective potential is
$V_{\mathrm{eff}}(0,m_2,0)$. The critical value $\delta_2=0$
corresponds to $M=m_2$.

    \item If $\Delta \delta=\alpha_1-\alpha_2=\delta_2-\delta_1<0$,
then $V_{\mathrm{eff}}(m_1,0,0)$ is the ``global'' minimum along the
\underline{$M=0$ line}, whereas if $\Delta \delta>0$ then the
``global'' minimum of the effective potential is
$V_{\mathrm{eff}}(0,m_2,0)$. The critical value $\Delta \delta=0$
corresponds to $m_1=m_2$.

\end{itemize}
Following an approach similar to the one described in the appendix
of Ref.~\cite{symmetric}, one can show that the effective potential
in the case when all three masses $m_i$ and $M$ are nonzero, has a
local minimum intermediate between the ``corner'' values. In
conclusion, the global minimum of the effective potential at
$T=\mu=0$ is always located at one of the corners of triangle
depicted in Fig.~\ref{triangle}. In Table~\ref{corners} we show
which corner corresponds to the global minimum of the effective, as
a function of the relative values of the parameters $\delta_1$
and~$\delta_2$.

\begin{widetext}

\section{Phase Structure}
\label{phase}

We discuss now the effective potential~\eqref{finite} corresponding
to the three possible vacuum phases identified in the previous
section. We have:
\begin{itemize}
   \item{$m_1=m_2=0$}.
In this case our model reduces to the pure Cooper-pairing
model~\cite{ref:paper1}, with $M^{*2} = \Delta^2$ is the dynamically
generated gap. Here, we choose $m_{10}=m_{20}=0$,
$M_0^2=\Delta^2=\gamma$, and $\bar \beta = 0$. Then, we can write
\begin{align}
   & V_{\mathrm{eff}}(m_i,M) = \
   m_1^2 \bar \alpha_1 + m_2^2 \bar \alpha_2
   \\ \notag &
   -
   \frac{1}{2}
   \int_{0}^{\infty} [\mathrm d k_1]
   \biggl \{
   k_+(m_+)
   + k_-(m_+)
   + k_+(m_-)
   + k_-(m_-)
   - 4 k_1
   -
   \frac{2(M^2 \!+m_1^2 \!+m_2^2)}{\sqrt{k_1^2 + \Delta^2}}
   \\ \notag &
   +\! \frac{2}{\beta} \ln\bigl [ 1\!+ \!e^{-\beta k_+(m_+)}\bigr ]
   \!+\! \frac{2}{\beta} \ln\bigl [ 1\!+ \!e^{-\beta k_-(m_+)}\bigr ]
   \!+\! \frac{2}{\beta} \ln\bigl [ 1\!+ \!e^{-\beta k_+(m_-)}\bigr ]
   \!+\! \frac{2}{\beta} \ln\bigl [ 1\!+ \!e^{-\beta k_-(m_-)}\bigr ]
   \biggr \}
   \>.
\end{align}
If we set $m_1=m_2=0$, then the chemical potential becomes
irrelevant and can be transformed away. We obtain
\begin{align}
   & V_{\mathrm{eff}}(m_i,M) = \
   \frac{M^2}{4\pi}
   \Bigl [ \ln \frac{M^2}{\Delta^2} - 1 \Bigr ]
   - \frac{2}{\beta}
   \int_{0}^{\infty} [\mathrm d k_1]
   \Bigl [
   \ln\bigl ( 1\!+ \!e^{-\beta E}\bigr )
   + \ln\bigl ( 1\!+ \!e^{-\beta E}\bigr )
   \Bigr ]
   \>,
\end{align}
with $E=\sqrt{ k_1^2 + m_2^2 }$.

The effective potential for the Cooper pair sector at finite
temperature is $\mu$ independent. The model has a second-order phase
transition to the unbroken phase at a critical
temperature~\cite{ref:paper1}, $T_\mathrm{c} = (\Delta/\pi) \,
e^{\gamma_E}$, where $\gamma_E=0.577\cdots$ is Euler's constant.

\item{$M=m_2=0$.}
This is the Gross-Neveu sector~\cite{ref:GN2}. Here, $m_1^{*2} =
m_{\mathrm{F}}^2$ is the dynamically generated fermion mass. We
choose $M_0=m_{20}=0$ and $m_{10}^2=m_{\mathrm{F}}^2=\gamma$.
Furthermore, we have $g_{1\mathrm{R}}^2=\pi$ and $\bar \alpha_1 =
0$. Thus, we can write
\begin{align}
   & V_{\mathrm{eff}}(m_i,M) = \
   M^2 \bar \beta + m_2^2 \bar \alpha_2
   \\ \notag &
   -
   \frac{1}{2}
   \int_{0}^{\infty} [\mathrm d k_1]
   \biggl \{
   k_+(m_+)
   + k_-(m_+)
   + k_+(m_-)
   + k_-(m_-)
   - 4 k_1
   -
   \frac{2(M^2 \!+m_1^2 \!+m_2^2)}{\sqrt{k_1^2 + m_{\mathrm{F}}^2}}
   \\ \notag &
   +\! \frac{2}{\beta} \ln\bigl [ 1\!+ \!e^{-\beta k_+(m_+)}\bigr ]
   \!+\! \frac{2}{\beta} \ln\bigl [ 1\!+ \!e^{-\beta k_-(m_+)}\bigr ]
   \!+\! \frac{2}{\beta} \ln\bigl [ 1\!+ \!e^{-\beta k_+(m_-)}\bigr ]
   \!+\! \frac{2}{\beta} \ln\bigl [ 1\!+ \!e^{-\beta k_-(m_-)}\bigr ]
   \biggr \}
   \>.
\end{align}
If we set $M=m_2=0$, then we obtain $m_+=m_-=m_1$, $k_\pm = E \pm
\mu$ with $E=\sqrt{ k_1^2 + m_1^2 }$, and the effective potential in
the Gross-Neveu sector can be written as
\begin{align}
   & V_{\mathrm{eff}}(m_i,M) = \
   \frac{m_1^2}{4\pi}
   \Bigl [ \ln \frac{m_1^2}{m_{\mathrm{F}}^2} - 1 \Bigr ]
   - \frac{2}{\beta}
   \int_{0}^{\infty} [\mathrm d k_1]
   \Bigl \{
   \ln\bigl [ 1\!+ \!e^{-\beta (E+\mu)}\bigr ]
   + \ln\bigl [ 1\!+ \!e^{-\beta (E-\mu)}\bigr ]
   \Bigr \}
   \>.
\end{align}
We note that the effective potential in the Gross-Neveu sector is
similar to the effective potential in the Cooper pair sector, with
$m_{\mathrm{F}}$ replacing $\Delta$, but depends on the chemical
potential $\mu$ at $T\neq 0$.

The Gross Neveu model has spontaneous symmetry breaking at zero
chemical potential and zero temperature. At zero temperature, the
model undergoes a first-order phase transition to the unbroken
symmetry phase as we increase the chemical
potential~\cite{ref:GN2,ref:minakata}. At zero chemical potential,
the symmetry is restored through a second order phase transition at
the critical temperature~\cite{symmetric}, $T_\mathrm{c} =
(m_{\mathrm{F}}/\pi) \, \exp \bigl [\gamma_E - 7 \mu^2 \zeta(3)/
(4\pi^2 T_\mathrm{c}^2) \bigr ]$. Thus, the phase diagram of the
Gross-Neveu model has a tricritical point, with approximate
values~\cite{ref:GN2} $\mu_c/m_F = 0.608,\, T_c/m_F = 0.318$.

\item{$m_1=M=0$.}
Here, $m_2^{*2} = \delta m^2$ is the dynamically generated mass
asymmetry. This case is identical with the Gross-Neveu sector case,
with $m_{\mathrm{F}}$ being replaced by $\delta m$: We choose
$m_{10}=M_0=0$, $m_{20}^2=\delta m^2=\gamma$, and
$g_{1\mathrm{R}}^2=\pi$ (or $\bar \alpha_2 = 0$). The effective
potential becomes
\begin{align}
   & V_{\mathrm{eff}}(m_i,M) = \
   M^2 \bar \beta + m_1^2 \bar \alpha_1
   \\ \notag &
   -
   \frac{1}{2}
   \int_{0}^{\infty} [\mathrm d k_1]
   \biggl \{
   k_+(m_+)
   + k_-(m_+)
   + k_+(m_-)
   + k_-(m_-)
   - 4 k_1
   -
   \frac{2(M^2 \!+m_1^2 \!+m_2^2)}{\sqrt{k_1^2 + \delta m^2}}
   \\ \notag &
   +\! \frac{2}{\beta} \ln\bigl [ 1\!+ \!e^{-\beta k_+(m_+)}\bigr ]
   \!+\! \frac{2}{\beta} \ln\bigl [ 1\!+ \!e^{-\beta k_-(m_+)}\bigr ]
   \!+\! \frac{2}{\beta} \ln\bigl [ 1\!+ \!e^{-\beta k_+(m_-)}\bigr ]
   \!+\! \frac{2}{\beta} \ln\bigl [ 1\!+ \!e^{-\beta k_-(m_-)}\bigr ]
   \biggr \}
   \>.
\end{align}
When $m_1=M=0$, we obtain
\begin{align}
   & V_{\mathrm{eff}}(m_i,M) = \
   \frac{m_2^2}{4\pi}
   \Bigl [ \ln \frac{m_2^2}{m_{\mathrm{F}}^2} - 1 \Bigr ]
   - \frac{2}{\beta}
   \int_{0}^{\infty} [\mathrm d k_1]
   \Bigl \{
   \ln\bigl [ 1\!+ \!e^{-\beta (E+\mu)}\bigr ]
   + \ln\bigl [ 1\!+ \!e^{-\beta (E-\mu)}\bigr ]
   \Bigr \}
   \>,
\end{align}
with $E=\sqrt{ k_1^2 + m_2^2 }$.

\end{itemize}
\end{widetext}


\section{Conclusions}
\label{concl}

In this paper we studied a generalization of a simple model with two
spontaneously broken symmetry phases first discussed in
Ref.~\cite{symmetric}. The model is studied in (1+1) dimensions
within the leading order in large N approximation.  We find  that the phase
diagram with two different masses is similar to the case when the
masses are the same in that depending on the values of the two
renormalized coupling constants one is always in one of three
distinct phases.

A generalization of this model to higher dimensions including the
case of spatially inhomogeneous order parameters is under
development. We expect that the results found here in 1+1 for
homogeneous order parameters will persist in mean field regardless
of the number of dimensions. Thus, in order to obtain phase
coexistence we will have to consider inhomogeneous condensates. For
this type of condensates we expect a small coexistence region
similar to the Larkin-Ovchinikov-Fulde-Ferrel state~\cite{loff} in
ferromagnetic superconductors, while preserving the tricritical
point.

After this work was completed, we became aware of a new study of the
massive Gross-Neveu model using semi-classical methods in the
large-N limit~\cite{massive}.  These semi-classical methods display
a kink-antikink crystal phase and it would be interesting to also
perform such an analysis for the model of this paper.  In 1+1
dimensions, semi-classical analyses such as those described by
Dashen, Hasslacher and Neveu~\cite{dhn} often gave exact answers for
the S-matrix (especially in exactly solvable models)  and include
more information than the straight forward study of the effective
potential in the large-N approximation.



\begin{thebibliography}{99}


\bibitem{raj}
   M.~Alford, K.~Rajagopal, and F.~Wilczek,
   Phys. Lett. B \textbf{422}, 247 (1998).

\bibitem{rapp}
   R.~Rapp, T.~Sch\"afer, E.~V.~Shuryak, and M.~Velkovsky,
   Phys. Rev. Lett. \textbf{81}, 53 (1998).

\bibitem{early}
   B.~K\"ampfer,
   Ann. Phys. (Leipzig) \textbf{9}, 606 (2000).

\bibitem{kbb1}
   N.~D. Mathur \emph{et al.}, Nature \textbf{394}, 39 (1998);
   S.~S. Saxena \emph{et al.}, \textit{ibid.} \textbf{406}, 587 (2000).

\bibitem{kbb2}
   J.~A. Hertz, Phys. Rev. B \textbf{14}, 1165 (1976);
   A.~J. MILLIS, Phys. Rev. B \textbf{48}, 7183 (1993).

   \bibitem{symmetric}
   A.~Chodos, F.~Cooper, W.~Mao, H.~Minakata, A.~Singh,
   Phys. Rev. D~\textbf{61}, 045011 (2000).

\bibitem{kbb3}
   N.~I. Karchev, K.~B. Blagoev, K.~S. Bedell, and P.~B. Littlewood,
   Phys. Rev. Lett. \textbf{86}, 846 (2001);
   K.~B. Blagoev, K.~S. Bedell, and P.~B. Littlewood
   \textit{ibid.} \textbf{92}, 199706 (2004).

\bibitem{loff}
   A.~I. Larkin and Yu.~N. Ovchinnikov, Zh. Eksp. Teor. Fiz.
   \textbf{47}, 1136 (1964) [Sov. Phys. JETP \textbf{20}, 762 (1975)];
   P.~Fulde and R.~A. Ferrell, Phys. Rev. \textbf{135}, A550 (1964).

\bibitem{raj2}
   J.~A. Bowers and K.~Rajagopal,
   Phys. Rev. D \textbf{66}, 065002 (2002);
   J.~Kundu and K.~Rajagopal, Phys. Rev. D \textbf{65}, 094022 (2002).

\bibitem{ref:GN}
D.~J. Gross and A.~Neveu, Phys. Rev. {\bf D10}, 3235 (1974).

   \bibitem{ref:paper1}
   A.~Chodos, H.~Minakata, and F.~Cooper, Phys. Lett.
   B~\textbf{449}, 260 (1999).

\bibitem{ref:super1}
D.~Bailin and A.~Love, Phys. Rep. {\bf 107}, 325 (1984); M.~Iwasaki
and T.~Iwado, Phys. Lett. B {\bf 350}, 163 (1995); M.~Alford,
K.~Rajagopal and F.~Wilczek, \textit{ibid.} {\bf 422}, 247 (1998).

\bibitem{ref:super2}
M.~Alford, K.~Rajagopal and F.~Wilczek, Nucl. Phys. {\bf A638}, 515c
(1998);  Nucl. Phys. {\bf B537}, 443 (1999); J.~Berges and
K.~Rajagopal, \textit{ibid.} {\bf B538}, 215 (1999); T.~Sch\"{a}fer,
Nucl. Phys. {\bf A642}, 45 (1998); T.~Sch\"{a}fer and F. Wilczek,
Phys. Rev. Lett. {\bf 82}, 3956 (1999).


\bibitem{ref:mermin}
S.~Coleman, Commun. Math. Phys. {\bf 31}, 259 (1973); N.~D. Mermin
and H.~Wagner, Phys. Rev. Lett. {\bf 17}, 1133 (1966).

\bibitem{ref:Witten}
E.~Witten, Nucl. Phys. {\bf B145}, 110 (1978).


\bibitem{ref:hub}
J.~Hubbard, Phys. Rev. Lett. {\bf 3}, 77 (1959); R.~L. Stratonovich,
Dokl. Akad. Nauk. SSSR {\bf 115}, 1097 (1957) [Sov. Phys. Dokl. {\bf
2}, 416 (1957)]; S.~Coleman, {\it Aspects of Symmetry} (Cambridge
University Press, Cambridge, England, 1985), p. 354.

\bibitem{lemma}
If
$
   \mathcal{M} =
   \Bigl [
   \begin{array}{cc}
   A & B \\
   C & D
   \end{array}
   \Bigr ]
$, then
$
   \ln \mathrm{det}(\mathcal{M})
   =
   \mathrm{Tr} \ln (A D)
   +
   \mathrm{Tr} \ln
   \bigl [
   \mathbf{1} -
   D^{-1}C \ A^{-1} B
   \bigr ]
$.

\bibitem{1std}
Note $\displaystyle{\delta_m} = 2 m \, \delta_{m^2}$,
$\displaystyle{\delta_B} = 4 B^\dag \, \delta_{M^2}$, and
$\displaystyle{\delta_{B^\dag}} = 4 B \, \delta_{M^2}$.

\bibitem{1+1s}
$A= (\mu^2 - k_0^2) + (m_1^2 - k_1^2) $, $\vec B= 2 \bigl ( m_1
\mu$, $- \mathrm{i}m_1 k_1$, $k_0 k_1
   \bigr )
$, $A'= (\mu^2 - k_0^2) - (m_1^2 - k_1^2) $, $\vec B' = - 2 \bigl (
m_1 k_0$, $\mathrm{i}m_1 k_1$, $\mu k_1
   \bigr )
$.

\bibitem{1+1a}
$A = (\mu^2 - k_0^2) + (m_1^2 - k_1^2) + m_2^2 $, $\vec B = 2 \bigl
( m_1 \mu$, $- \mathrm{i}m_1 k_1$, $k_0 k_1
   \bigr )
$,
$C = 2 m_1 m_2
$,
$\vec D = 2 m_2 \bigl ( \mu$, $- \mathrm{i} k_1$, $0
   \bigr )
$, and $A' = (\mu^2 - k_0^2) - (m_1^2 - k_1^2) - m_2^2 $, $\vec B' =
- 2 \bigl ( m_1 k_0$, $\mathrm{i}m_1 k_1$, $\mu k_1
   \bigr )
$, $C' = - 2 m_1 m_2 $, $\vec D' = - 2 m_2 \bigl ( k_0$,
$\mathrm{i}k_1$, $0
   \bigr )
$.

\bibitem{int1}
$\displaystyle{\int_0^\lambda \!\! \mathrm{d}k \bigl ( \sqrt{k^2 +
m^2} - k \bigr ) \rightarrow \frac{m^2}{2}} \Bigl [ \frac{1}{2} +
\ln \Bigl ( \frac{2\Lambda}{|m|} \Bigr ) \Bigr ]$, for
large~$\Lambda$.

\bibitem{2ndd}
Here, we have used ${\frac{1}{4} \, \delta_B \delta_{B^\dag} =
\delta_{M^2} + M^2 \, \delta_{M^2} \delta_{M^2}}$, and ${\frac{1}{2}
\, \delta^2_{m^2} = \delta_{m^2} + m \delta_m \delta_{m^2}}$.

\bibitem{lim_M0}
For $M_0 \rightarrow 0$, we have $\displaystyle{\frac{1}{8\pi}
\frac{m_0}{M_0} \ln \Bigl | \frac{M_0 - m_0}{M_0 + m_0} \Bigr |} = -
\frac{1}{4\pi}$.

\bibitem{ref:GN2}
U.~Wolff, Phys. Lett. B {\bf157}, 303 (1985); L.~Jacobs, Phys. Rev.
{\bf D10}, 3956 (1974); B.~Harrington and A.~Yildiz,  Phys. Rev.
{\bf D11}, 779 (1975).

\bibitem{ref:minakata}
A.~Chodos and H.~Minakata, Phys. Lett. A {\bf191}, 39 (1994); Nucl.
Phys. {\bf B490}, 687 (1997).

\bibitem{massive}
O.~Schnetz, M.~Thies, and K.~Urlichs, hep-th/0511206. See also
M.~Thies and K.~Urlichs, Phys. Rev. D \textbf{67}, 125015 (2003);
M.~Thies, \textit{ibid.} \textbf{69}, 067703 (2004); O.~Schnetz,
M.~Thies, and K.~Urlichs, Ann. Phys. \textbf{314}, 425 (2004).

\bibitem{dhn}
R.F.~Dashen, B.~Hasslacher, and A.Neveu, Phys. Rev. D \textbf{12},
2443 (1975).


\end{thebibliography}
\end{document}